\documentclass[]{article}

\usepackage{graphicx}
\usepackage{epstopdf}

\usepackage{amssymb}
\usepackage{latexsym}

\usepackage[
    centering,
    marginparwidth=0in,
    textwidth=6.5in,
    marginparsep=2em,
    top=2.5cm,
    bottom=2.5cm,
]{geometry}

\usepackage{tensor}
\ifpdf
\DeclareGraphicsExtensions{.eps,.pdf,.png,.jpg}
\else
\DeclareGraphicsExtensions{.eps}
\fi
\usepackage{amsmath}
\usepackage{amssymb}

\usepackage{tikz}
\usepackage{tikz-cd}
\usepackage{pgfplots}

\newcommand{\ie}{i.\,e.}%
\newcommand{\eg}{e.\,g.}%
\newcommand{\st}{s.\,t.}%
\newcommand{\wrt}{w.r.t.}%

\newcommand{\formComma}{\,\text{,}}
\newcommand{\formPeriod}{\,\text{.}}

\newcommand{\energy}{\mathcal{F}}

\newcommand{\landau}{\mathcal{O}}

\newcommand{\surf}{\mathcal{S}}
\newcommand{\surfh}{\surf_{h}}

\newcommand{\normal}{\boldsymbol{\nu}}

\newcommand{\tangent}{\textup{T}}

\newcommand{\dir}{\boldsymbol{P}}
\newcommand{\dirTwo}{\boldsymbol{M}}
\newcommand{\dirSurf}{\boldsymbol{p}}
\newcommand{\dirTwoSurf}{\boldsymbol{m}}

\newcommand{\gb}{\boldsymbol{g}}%
\newcommand{\Gb}{\boldsymbol{G}}%
\newcommand{\tb}{\boldsymbol{t}}%
\newcommand{\qb}{\boldsymbol{q}}%
\newcommand{\Qb}{\boldsymbol{Q}}%
\newcommand{\Psib}{\boldsymbol{\Psi}}%
\newcommand{\psib}{\boldsymbol{\psi}}%

\newcommand{\Pb}{\boldsymbol{P}}%

\newcommand{\dV}{\text{d}V}%
\newcommand{\dS}{\text{d}\surf}%
\newcommand{\dxi}{\text{d}\xi}%

\newcommand*{\rom}[1]{\textup{\uppercase\expandafter{\romannumeral#1}}}

\newcommand{\shapeOperator}{\boldsymbol{B}}%
\newcommand{\gaussianCurvature}{\mathcal{K}}%
\newcommand{\meanCurvature}{\mathcal{H}}%

\DeclareRobustCommand{\GGamma}{\text{\raisebox{\depth}{\scalebox{1}[-1]{$\mathbb{L}$}}}}

\newcommand{\Laplace}{\boldsymbol{\triangle}}
\newcommand{\laplace}{\triangle}

\newcommand{\trace}{\operatorname{tr}_2}%
\newcommand{\traceR}{\operatorname{tr}_3}%

\newcommand{\Proj}[1]{\operatorname{\Pi}_{#1}}

\newcommand{\ProjQ}{\Proj{\mathcal{Q}}}

\newcommand{\meanc}{\mathcal{H}}

\newcommand{\atsurf}[1]{\left. #1 \right|_{\surf}}

\title{\textbf{Properties of surface Landau-de Gennes Q-tensor models}}
\author{Michael Nestler,$^{\ast}$\textit{$^{a}$} Ingo Nitschke,\textit{$^{a}$}, Hartmut L\"owen\textit{$^{b}$} and Axel Voigt\textit{$^{a,c,d,e}$}}

\begin{document}

\maketitle

\footnotetext{\textit{ $^{\ast}$~Corresponding author: E-mail: michael.nestler@tu-dresden.de}}
\footnotetext{\textit{ $^{a}$~Institut f\"ur Wissenschaftliches Rechnen, Technische	Universit\"at Dresden, 01062 Dresden, Germany }}
\footnotetext{\textit{ $^{b}$~Institut f\"ur Theoretische Physik II - Soft Matter, Heinrich-Heine-Universit\"at D\"usseldorf, 40225 D\"usseldorf, Germany }}
\footnotetext{\textit{ $^{c}$~Dresden Center for Computational Materials Science (DCMS), Technische	Universit\"at Dresden, 01062 Dresden, Germany }}
\footnotetext{\textit{ $^{d}$~Center for Systems Biology Dresden (CSBD), Pfotenhauerstr. 108, 01307 Dresden, Germany }}
\footnotetext{\textit{ $^{e}$~Cluster of Excellence Physics of Life, TU Dresden, 01062 Dresden, Germany }}

\begin{abstract}
Uniaxial nematic liquid crystals whose molecular orientation is subjected to a tangential anchoring on a curved surface offer a non trivial interplay between the geometry and the topology of the surface and the orientational degree of freedom. We consider a general thin film limit of a Landau-de Gennes Q-tensor model which retains the characteristics of the 3D model. From this, previously proposed surface models follow as special cases. We compare fundamental properties, such as alignment of the orientational degrees of freedom with principle curvature lines, order parameter symmetry and phase transition type for these models, and suggest experiments to identify proper model assumptions.
\end{abstract}

\section{Introduction}

Liquid crystals \cite{de_Gennes_book,Chaikin_Lubensky} consist of particles that possess both translational and
orientational degrees of freedom. If these particles are constrained to the tangent bundle of a curved surface interesting phenomena emerge, which result from the tight coupling of the elastic and bulk free energies of the liquid
crystal with topological and geometrical properties of the surface. There are various experimental realization  \cite{shells1,Leon,Liang,Lagerwall_colloidal,Stannarius1,Lagerwall,Jia1,Xu,Jia2,Aarts1,Chaikin_2018} and particle-based computer simulations \cite{Dzubiella,Zannoni,Shin_PRL_2008,Dhakal_2012,Koning,van_der_Schoot_PRE,Assoian,Stark,Prinsen,Liublana,Geigenfeind,Yeomans,Henkes,Alaimo,Allahyarovetal_SM_2017,Sittaetal_PCCP_2018,Allahyarovetal_SM_2018}, which mainly focus on the emergence and position of topological defects on spherical or more complex surfaces. However, not only defects are tightly linked to topological and geometrical properties of the surface, also other fundamental issues, such as alignment of the orientational degree of freedom with principle curvature lines, order parameter symmetries, phase transition type and curvature induced phase transitions are of fundamental interest, but are much less explored. We will address these issues for uniaxial nematic liquid crystals within a field-theoretical desciption of a surface Landau-de Gennes Q-tensor model. Various such models have been proposed  \cite{Kraljetal_SM_2011,Napolietal_PRE_2012,Napolietal_PRL_2012,Jeseneketal_SM_2015,golovaty2015dimension,nitschke2018nematic}. They strongly differ in the coupling mechanism between orientational ordering of the nematic liquid crystal and the geometric properties of the surface. These coupling terms strongly depend on the made assumptions in the derivation and, as will be shown, have strong consequences on the fundamental properties of phase transition type and order parameter symmetries. 

Liquid crystals on curved surfaces are somehow in between 2D and 3D and could thus show properties of both dimensions. Lets first compare properties of the established Landau-de Gennes Q-tensor theory for nematic liquid crystals in 2D and 3D \cite{de_Gennes_book}: The nematic phase can be stable in both dimensions. However, already the isotropic-to-nematic phase transition qualitatively differs between 2D and 3D. While it is of first-order in 3D, it is controversially discussed if this generally holds in 2D. Even if it can be proven that first-order isotropic-to-nematic phase transitions are possible in 2D  \cite{vanEnteretal_PRL_2002,vanEnteretal_JPA_2006}, computer simulations and experiments indicate qualitatively different behaviour including continuous, first order and even absence of phase transitions, see e.g. \cite{Wensinketal_JPCM_2007,Vink_PRE_2014,Wittmannetal_JCP_2017}. In \cite{Fishetal_PRE_2010} an overview of these theoretical arguments, computer simulations and experimental observations for thin films and 2D systems is provided. The situation on curved surfaces should somehow reflect these properties. However, in some of the proposed surface models first order phase transitions are not possible. Another aspect highlighting the differences is discussed in \cite{majumdar2010equilibrium} by comparing mean field theories in 2D \cite{straley1971liquid} and 3D \cite{stephen1974physics}. The definitions yield different eigenvalue spectra in the order tensor, which corresponds to either a symmetry under in plane rotations by $90^{\circ}$ in 2D or a rotational symmetry (\wrt\ the average particle direction) in 3D. How a nematic liquid crystal on a curved surface fits into this picture is open. Some of the proposed surface models yield an eigenvalue spectrum as in 2D, others as in 3D. The third aspect considers the mean orientation. If not induced by external fields or boundary conditions the mean orientation in nematic liquid crystals is arbitrary in 2D and 3D. This changes on curved surfaces, where it should be preferential to align with the principle curvature lines. This aspect has been discussed previously and is identified with the influence of extrinsic curvature terms \cite{Napolietal_PRE_2012,Napolietal_PRL_2012} in surface models, which again are considered in some but not all of the proposed models.

We will review the proposed surface Landau-de Gennes Q-tensor models under the aspect of these fundamental properties. Furthermore we propose a version of a surface Landau-de Gennes Q-tensor model which effectively describes surface liquid crystals retaining the 3D phase transition type and eigenvalue spectra. The previously proposed models follow as special cases and we discuss under which assumptions fundamental properties get lost. The paper is structured as follows. In Section \ref{sec:ThinFilmLimit} we briefly review the 3D Landau-de Gennes Q-tensor model \cite{de_Gennes_book} and propose its thin film limit under generic anchoring conditions. In Section \ref{sec:FlatThinFilm} this model together with its special cases is discussed in a flat 2D scenario with respect to their fundamental properties. We demonstrate the general model to retain the 3D properties. With this established, we apply the models to curved surfaces and discuss the effects of curvature on the ordering of the liquid crystal in Section \ref{sec:CurvedThinFilm}. We summarize our findings and discuss them in a general framework in Section \ref{sec:Conclusion}. As all these results do not depend on specific material parameters, we only consider a one-constant approximation. Details on derivations and used numeric methods can be found in the Appendix.

\section{Thin Film Limit of Q-Tensor Model}
\label{sec:ThinFilmLimit}
In this section we present essential notions and properties of the Landau-de Gennes Q-tensor model in 3D \cite{de_Gennes_book} and derive a generic surface model as a thin film limit. Restrictions to enforce uniaxiality and special cases, which link the model to previously proposed surface Landau-de Gennes Q-tensor models follow. 

\paragraph*{Q-Tensor Model in 3D} We consider rod like particles with a head tail symmetry in a volume $V\subset \mathbb{R}^3$. The symmetric and trace free tensorial order parameter $\Qb$ is defined by 
\begin{align}
\Qb &= S\left(\dir\otimes\dir - \frac{1}{3} \Gb\right), \quad \sigma(\Qb) = \pm \left[\frac{2}{3},\, -\frac{1}{3},\, -\frac{1}{3}\right]S \label{eq:QtensorDefinition}
\end{align}
where $\dir$ denotes the principal director, defined by the average orientation of the particles, $S$ the scalar order parameter, encoding the degree of alignment by the particles with the average direction, and $\Gb$ the metric of $V$. $\sigma(\Qb)$ denotes the uniaxial eigenvalue spectra. The phase of prevalent liquid like material properties is characterized by an isotropic ordering of particles and $S=0$. In the nematic phase the particles tend to preferentially align with the average direction and $S\rightarrow S^* > 0$. The Landau-de Gennes Q-tensor model is based on the free energy 
\begin{align}
\energy(\Qb)  =&  \int_V  \frac{L}{2}\|\nabla \Qb\|^2 + \omega \left[ a \traceR \Qb^2 +\frac{2}{3}b \traceR \Qb^3 + c \traceR \Qb^4 \right] \,\mathrm{d}V. \label{eq:BulkEnergyOneConst}
\end{align} 
The first term, the elastic energy, penalizes any spatial deviations from the ground state. For sake of simplicity we consider a one-constant approximation with elastic parameter $L$. For more general models see \cite{Schieleetal_PSS_1983,Berremanetal_PRA_1984,Lubensky_PRA_1970,Longaetal_LC_1987,Balletal_MCLC_2010} and for a corresponding surface model \cite{nitschke2018nematic}. In the remaining terms, the state potential, we have factored out $\omega > 0$ such that classic phenomenological constants are given by $A= \omega\, a$, $B= \omega\, b$ and $C= \omega\, c$. We use the trace notion given by $\traceR \Qb = \Qb : \Gb$ as a full contraction with the space metric which coincides with the Frobenius tensor norm by $\traceR \Qb^2 = \|\Qb\|^2 = \sum_{ij} Q_{ij}^2$. The state potential can be expressed in terms of $S$ such that the choice of $a,\,b$ and $c$ define the preferred ordering $S^*$ as local minima of the state potential.\\
Following \cite{majumdar2010equilibrium,Balletal_MCLC_2010} we point out that the tensorial order parameter $\Qb$ in this model is not restricted to the uniaxial eigenvalue spectra $\sigma(\Qb)$ as defined in \eqref{eq:QtensorDefinition}. Spectra beside the uniaxial configuration are called biaxial and can be related to different liquid crystal systems. To track such situations, a biaxiality measure is introduced \cite{Kaiseretal_JNET_1992}
\begin{align}
U(\Qb) = 1 - 6 \frac{(\traceR \Qb^3)^2}{(\traceR \Qb^2)^3 }  
\label{eq:BiaxialityMeassure}
\end{align}
for which $U(\Qb)=0$ if and only if $\Qb$ is uniaxial. This measure is discussed in detail in \cite{majumdar2010equilibrium} and it is established that for $b<0$ state potential minimas $\Qb^*$ are uniaxial.

\paragraph*{Planar Anchoring and Tensor Decomposition} For the boundary of the volume $V$, $\partial V$, with outward pointing normal $\normal$ planar anchoring of uniaxial Q-tensors is modeled by a bare surface energy as discussed in \cite{golovaty2015dimension},
\begin{align}
\energy_B(\Qb) = \int_{\partial V} \!\alpha \left[ \normal \cdot \Qb \cdot \normal - \beta \right]^2 + \gamma \| \left(\mathbb{I}_3 - \normal \otimes \normal\right)\cdot \left(\Qb \cdot \normal\right) \|^2 \, \mathrm{d}\partial V,
\label{eq:anchoring}
\end{align}
with coefficients $\alpha$ and $\gamma$. The formulation can be interpreted as a penalty energy enforcing $\normal$ as an eigenvector of $\Qb$ with eigenvalue $\beta$. In the case of a tangential aligned uniaxial Q-tensor ($\dir \cdot \normal = 0$) this would translate into $\beta = -1/3 S$. Given this concept of prescribing a specific eigenvalue in boundary normal direction motivates the separation of $\Qb$ in a normal part $\beta$ and tangential parts $\qb$ ($\qb \cdot \normal = \normal \cdot \qb = 0$). We thereby choose a separation such that $\qb$ is symmetric and trace free (in the boundary domain sense)
\begin{align}
\Qb(\qb,\beta) = \qb - \frac{\beta}{2}\gb + \beta \normal\otimes\normal, \label{eq:qtensorDecomposition}
\end{align}
where $\gb$ is the metric of $\partial V$ and $\trace \qb = \qb : \gb = 0$ the associated notion of trace. Within this decomposition we can consider $\qb$ as a two dimensional Q-tensor on $\partial V$ with tangential principal director $\dirSurf$ ($\dirSurf \cdot \normal =0$)
\begin{align}
\qb = S \left(\dirSurf\otimes\dirSurf - \frac{1}{2}\gb\right), \quad \sigma(\qb) = \pm \left[\frac{1}{2},\, -\frac{1}{2},\, 0\right]S.
\end{align}
%Please note such notion of a two dimensional Q-tensor $\qb$ matches the model of \cite{Kralj2011} and $\traceR \qb^3 = \mathrm{det}\, \qb = 0$.

\paragraph*{Thin Film Limit Models}
Lets consider $V = \surfh$ as a thin film with thickness $h$, such that $\surfh = \surf \times [-h/2,h/2]$ and $\surf$ a regular surface. We perform the thin film limit $\lim_{h \rightarrow 0} 1/h \, \energy(\Qb) = \energy^{\surf}(\Qb)$ in the spirit of \cite{nitschke2018nematic} under boundary conditions $\normal \cdot \Qb \cdot \normal -\beta =0, \; \left(\mathbb{I}_3 - \normal \otimes \normal\right)\cdot \left(\Qb \cdot \normal\right) = 0$. For details see Appendix. By averaging out the normal direction we yield a surface model. Inserting \eqref{eq:qtensorDecomposition} in \eqref{eq:BulkEnergyOneConst} leads to
\begin{align}
\energy^{\surf}[\qb,\beta] = & \frac{1}{2}\int_{\surf} L \| \nabla \qb \|^2 + \frac{3}{2}L \| \nabla \beta \|^2  \nonumber \\ 
& \quad\quad- 6 L \meanCurvature \beta \langle \shapeOperator, \qb \rangle  +  L\| \shapeOperator\| ^2 \left( \trace \qb^2 + \frac{9}{2}\beta^2\right)\,\mathrm{d}\surf \nonumber 
\end{align}
\begin{align}
&+ \omega\int_{\surf}\frac{1}{2}\left(2a -2b\beta + 3c\beta^2\right) \trace \qb^2 + c \trace \qb^4 \nonumber \\
& \quad\quad + \frac{\beta^2}{8}\left(12a +4b\beta +9 c \beta^2\right)  \,\mathrm{d}\surf.
\label{eq:CurvedSurfaceBetaExplicitEnergy}
\end{align}
In contrast to \eqref{eq:BulkEnergyOneConst} all operators are defined by the Levi-Civita connection and inner products are considered at the surface. As in 3D we identify the first integral with the elastic energy and the second with the state potential. The first integral contains additional coupling terms, where $\shapeOperator$ denotes the shape operator and $\meanCurvature$ the mean curvature of $\surf$, for details see Appendix. These are extrinsic curvature contributions, the term $6 \meanCurvature \beta \langle \shapeOperator, \qb \rangle$ induces an alignment of $\dirSurf$ with one of the lines of principle curvatures depending on sign of $\meanCurvature$. The second term $\| \shapeOperator\| ^2 ( \trace \qb^2 + \frac{9}{2}\beta^2)$ poses an isotropic coupling between curvature and ordering. The term is closely related to the state potential such that curvature can locally deform the potential and can induce a phase transition. %Such effects have been discussed for the special case $\beta = -1/3 S^*$ in \cite{nitschke2018nematic}.

As evolution law we propose $L^2$-gradient flows of the energy \eqref{eq:CurvedSurfaceBetaExplicitEnergy} with independent variables $\qb$ and $\beta$, which read
\begin{align}
\partial_t \qb &= L \Laplace^{DG}_\surf \qb - ( L(\meanCurvature^2 - 2 \gaussianCurvature) - \omega(2 a -2b \beta + c(3 \beta^2 + 2 \trace \qb^2)))\qb \nonumber \\
& \qquad + 3 L \meanCurvature \beta \left(\shapeOperator - \frac{1}{2} \meanCurvature \gb \right) \label{eq:q}\\
\partial_t \beta &= L \laplace_\surf \beta - \omega (3c \beta^3 + b \beta^2) - \omega(2a + 2c \trace \qb^2) \beta \nonumber \\
& \qquad - 3L(\meanCurvature^2 - 2 \gaussianCurvature) \beta + 2 L \meanCurvature \langle \shapeOperator, \qb \rangle + \frac{2}{3} b \trace \qb^2 \label{eq:beta}
\end{align}
where $\Laplace^{DG}_\surf$ denote the surface Div-Grad (Bochner) Laplace operator, $\laplace_\surf$ the Laplace-Beltrami operator and $\gaussianCurvature$ the Gaussian curvature. Details of the derivation can be found in Appendix. Eqs. \eqref{eq:q} and \eqref{eq:beta} provide a general surface Landau-de Gennes Q-tensor model with a minimum of a priori assumptions. Due to this generality we expect, as in 3D, the solution space for the tensorial order parameter $\Qb(\qb,\beta)$ not to be restricted to the uniaxial eigenvalue spectra
$\sigma(Q)$ as defined in \eqref{eq:QtensorDefinition}. As the curvature terms can also locally influence the state potential, see discussion above, we expect a simple criterion such as $b < 0$, which enforces uniaxiality in 3D, not to hold on surfaces. To demonstrate this we solve \eqref{eq:q} and \eqref{eq:beta} numerically, see Appendix. Fig. \ref{fig:UniaxialityFlat}-A,B shows the equilibrium solution on a flat (2D) and a curved surface, respectively, demonstrating the distortion of uniaxiality by curvature.

\begin{figure}[ht!]
	\begin{center}
		\includegraphics[width=0.5\linewidth]{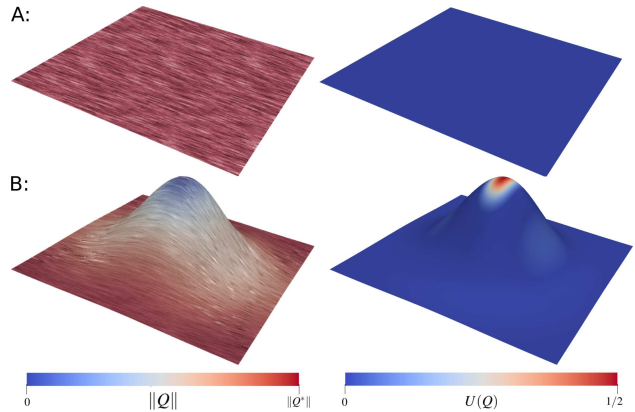}
	\end{center}
	\caption{\textbf{Curvature distorts Uniaxiality:} [A] In 2D energetic minima of unconstrained surface Q-tensors $\energy(\Qb(\qb,\beta))$ are uniaxial (for $b<0$, see \eg\ \cite{majumdar2010equilibrium}). (left) 2D yields uniform alignment of principal director (lines, obtained by line integral convolution (LIC) and constant $\|\Qb\|$ (colorscale). (right) Uniform values of biaxiality measure $U(\Qb) \approx 0$. [B] Curvature in domain distorts $\|\Qb\|$ and also uniaxiality. (left) Norm deficiency in high curvature areas. (right) Uniaxiality is substantially violated in high curvature regions where $U(\Qb) \approx 1/2$.}
	\label{fig:UniaxialityFlat}
\end{figure}

Instead of searching for a generalized criterion to guarantee uniaxiality on surfaces, which would have to include curvature effects, we use the criteria in 3D. We insert the decomposition \eqref{eq:qtensorDecomposition} in the biaxiality measure \eqref{eq:BiaxialityMeassure}, which leads a condition for $\beta$ enforcing uniaxial symmetry/eigenvalue spectra of Q-tensors, for details see Appendix. It reads
\begin{align}
\beta = \pm \frac{\sqrt{2}}{3} \| \qb \|. \label{eq:surfaceUniaxiality}
\end{align}
Using this constraint in \eqref{eq:CurvedSurfaceBetaExplicitEnergy} to eliminate $\beta$ leads to numerically cumbersome terms in the first variation, which is not further pursued. We instead add a penalty term to  \eqref{eq:CurvedSurfaceBetaExplicitEnergy} to enforce the constraint weakly,
\begin{align}
\omega_\beta \int_\surf \frac{1}{4} \left(\beta^2 - \frac{2}{9} \|\qb\|^2\right)^2 d\surf, \label{eq:surfaceUniaxialityPenalty}
\end{align}
with $\omega_\beta > 0$. This leads to numerically suitable additional terms to be added in \eqref{eq:q} and \eqref{eq:beta}, see Appendix. 
Previously proposed models consider special choices for $\beta$, which simplify \eqref{eq:q}. $\beta = 0$ yields a model with degenerate Q-tensors $\Qb = \qb$ as in \cite{Kraljetal_SM_2011,Napolietal_PRE_2012,Napolietal_PRL_2012,Jeseneketal_SM_2015}, for $\beta = -1/3 S^*$ we yield the model of approximate uniaxial Q-tensors \cite{nitschke2018nematic,Nitschkeetal_PRF_2019,Nitschkeetal_arXiv_2019}. In the following we will compare these three models.

\begin{figure*}[ht!]
	\begin{center}
		\includegraphics[width=0.85\linewidth]{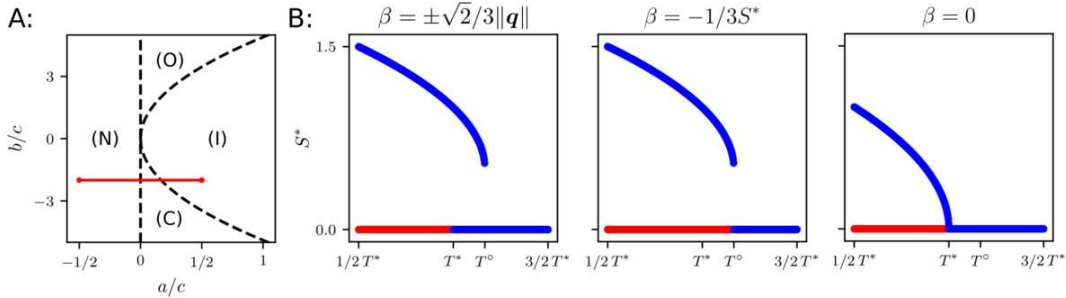}
	\end{center}
	\caption{\textbf{Types of Phase Transition in Surface Models:} [A] Phase portrait for bulk Landau-de Gennes Q-tensor model as defined in \eqref{eq:BulkEnergyOneConst}. Four distinct phases \wrt\ state potential parameters $a,\,b$ and $c$ exist. Only the isotropic phase is stable (I), only the nematic phase is stable (N) isotropic-to-nematic phase coexistence (C) and orthogonal ordering (O). Red line indicates a typical phase transition for $a(T) = \alpha(T-T^*),\, T \in [1/2T^*,3/2T^*]$ where $T^*$ denotes critical temperature such that isotropic phase loses stability, (I)-to-(N) transition. Here, the temperature $T^{\circ}$ marks the transition from only isotropic stable regime to coexistence, (I)-(C) transition. (O) denotes also a region of coexistence but due to $b/c >0$ a preferred ordering $S^*<0$ is observed. Such situations of orthogonal ordering (\wrt\ $\dir$) is not discussed here. [B] Stability of minima $S=\{0,S^*\}$ of state potential \wrt\ to temperature for $a(T)$ and choice of $\beta$ model. Blue lines denote stable minima while red lines indicate unstable minima.} %For $\beta = \sqrt{2}/3 \|\qb\|$(left) we observe first order type of phase transition at $T^{\circ}$ with distinct domain of phase coexistence for $T \in [T^*,T^{\circ}]$. For fixed $\beta = -1/3 S^*$(middle) we yield a first order phase transition at $T^{\circ}$ but no phase coexistence. With $\beta=0$(right) model we observe second order phase transition at $T^*$.
	\label{fig:PhaseTransitionTypesFlat}
\end{figure*}

\begin{figure*}[ht!]
	\begin{center}
		\includegraphics[width=0.85\linewidth]{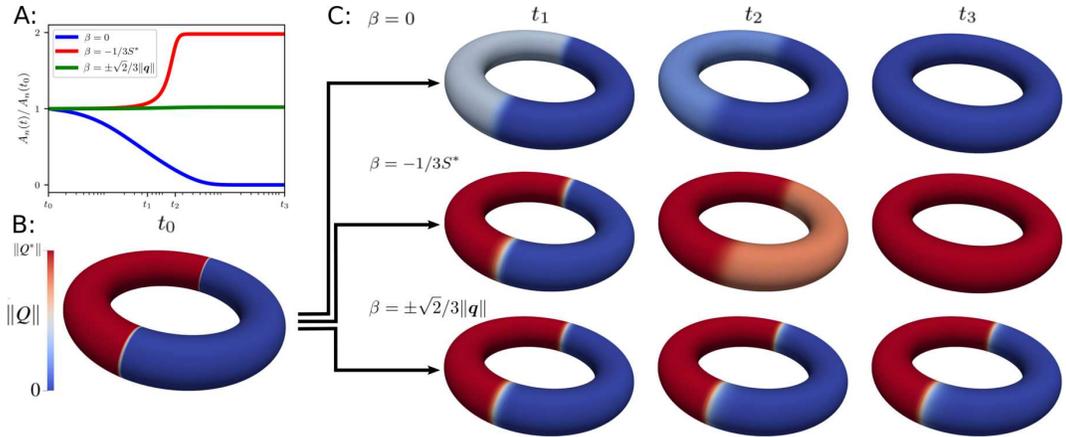}
	\end{center}
	\caption{\textbf{Phase Coexistence in Surface Models:} [A] Temporal evolution of area covered by nematic domain $A_{n}(t)$ vs initial nematic domain $A_{n}(t_0)$. Starting at equally sized domains for $\beta=0$ the nematic domain slowly shrinks, for $\beta = -1/3S^*$ the isotropic phase rapidly changes to nematic ordering, and for $\beta=\pm\sqrt{2}/3 \|\qb\|$ the ratio remains almost constant. [B] Initial distribution of isotropic (blue) and nematic (red) domains on torus ($R=2,\, r=1/2$). [C] Snapshots of $\|\Qb\| \in [0, \|\Qb^*\|]$ for $t_i \in [1e-4,1]$. Parameters: $L=1$, $a=1/4,\, b=-4,\, c=1$  and $\omega = 100$, $\omega_{\beta} =10$, $\omega_{\normal}=1000$.}
	\label{fig:betaModelPhaseCoexistence}
\end{figure*}

\section{Fundamental Properties of the Surface Models}
\label{sec:FlatThinFilm}

To keep the discussion as simple as possible we start with planar surfaces.

\paragraph*{Eigenvalue Spectra and Symmetries}
We recall the results of the mean field modeling approach for 3D systems, see \eg\ \cite{stephen1974physics},
\begin{align}
\Qb^{MF} = \int_{S^2} \left(\dirTwo\otimes\dirTwo - \frac{1}{3} \mathbb{I}_3\right) \phi_3(\dirTwo) \,\mathrm{d}\dirTwo, \quad \dirTwo \in \mathbb{R}^3,\; \|\dirTwo \| = 1
\end{align}
with $\dirTwo$ as particle orientation and $\phi_3$ the associated probability distribution. For planar alignment of $\dirTwo$ \wrt\ to the tangential direction, of a thin film, the mean field model implies an eigenvalue spectrum $\sigma(\Qb^{MF}) = [2/3, -1/3, -1/3]$. The mean field model in 2D, \eg\ \cite{straley1971liquid}, reads similar
\begin{align}
\qb^{MF} = \int_{S^1} \left(\dirTwoSurf\otimes\dirTwoSurf - \frac{1}{2} \mathbb{I}_2\right) \phi_2(\dirTwoSurf) \,\mathrm{d}\dirTwoSurf, \quad \dirTwoSurf \in \mathbb{R}^2,\; \|\dirTwoSurf \| = 1
\end{align}
but implies an eigenvalue spectrum of $\sigma(\qb^{MF}) = [1/2, -1/2]$. Considering the eigenvalue spectrum for the decomposed Q tensor \eqref{eq:qtensorDecomposition}
\begin{align}
\sigma(\Qb(\qb,\beta)) = \pm\left[ \frac{1}{2}\left(S- \beta \right),\; -\frac{1}{2}\left(S + \beta\right),\, \beta\right] \label{eq:EigenvalueSpectraQbeta}
\end{align}
we observe for $\beta=0$ a spectrum compatible with the 2D mean field theory, while $\beta=-1/3 S^*$ and $\beta = \pm\sqrt{2}/3\|\qb\|$ conform to 3D theory.\\
%Please note that these eigenvalue spectra correspond to symmetries in the local ordering which results in invariance of the Q-tensor field under associated transformations. While the 2D spectra implies the invariance of $\Qb(\qb, 0)$ under in plane rotations by $90^{\circ}$ the 3D Q-tensor \eg\ $\Qb(\qb,-\sqrt{2}/3\|\qb\|)$ is invariant under arbitrary rotation around the principal director $\dir$. Later situation only makes sense if we consider the surface model as an effective description of a thin film.

\paragraph*{Phase Transition Type}
We now turn to the isotropic-to-nematic phase transition. In the Landau-de Gennes Q-tensor model such transitions can be accounted for by a temperature dependent coefficient $a = a(T) = a_0(T-T^*)$ where $T^*$ denotes the critical temperature where isotropic phase is stable for $T>T^*$. Fig.\ref{fig:PhaseTransitionTypesFlat}-A shows the phase portrait of the Landau-de Gennes Q-tensor model for the 3D case \wrt\ $a/c,b/c$ with a typical transient for increasing $T$ from pure nematic phase, via phase coexistence of nematic and isotropic phase to pure isotropic phase. $T^{\circ}$ denotes the critical temperature where the nematic phase ceases to exist. In this framework the transition is discontinuous/first order.

We transfer this investigations of the transient from 3D to the surface models. For this purpose we consider states of uniform $\Qb(\qb,\beta)$ and evaluate the minima of the state potential contribution in \eqref{eq:CurvedSurfaceBetaExplicitEnergy} \wrt\ to $a(T)$ and the choice of $\beta$. In Fig. \ref{fig:PhaseTransitionTypesFlat}-B we have plotted the minima and their stability. Reviewing the results for $\beta = \pm\sqrt{2}/3\|\qb\|$ we observe a behavior identical to the 3D case. This is quite natural, since inserting the Q-tensor decomposition in 3D state potential energy density yields directly the surface counterpart. Therefore this surface model exhibits a first order phase transition type and enables phase coexistence. For the model of fixed $\beta = -1/3 S^*$ we observe a first order transition at $T^{\circ}$ as in the previous model but no phase coexistence for $T \in [T^*,T^{\circ}]$. In the case of $\beta = 0$ the transition type changes to continuous/second order and shifts to the lower temperature $T^*$.

\begin{figure*}[ht!]
	\begin{center}
		\includegraphics[width=0.95\linewidth]{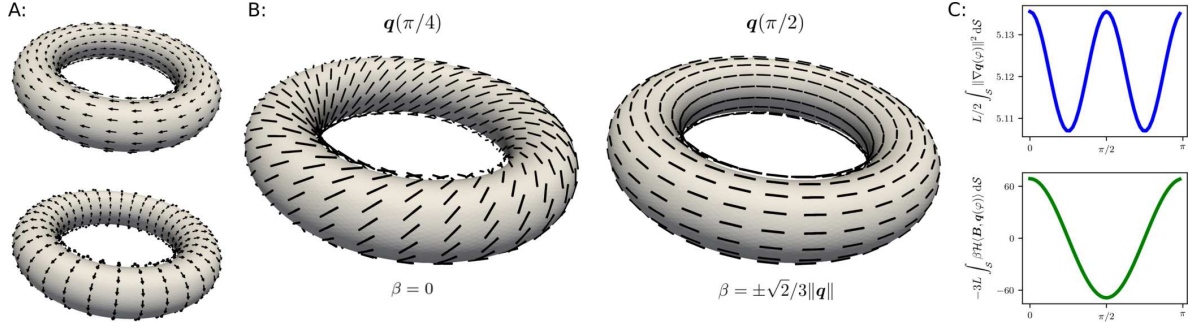}
	\end{center}
	\caption{\textbf{Distortion energy of defect free configurations of Torus:}[A] Director modes on torus ($R=2$, $r=1/2$).  $\dirSurf_R$(top) and $\dirSurf_r$(bottom) [B] Equilibrium states of evolution equations for $\beta=0$(left) and $\beta=\pm\sqrt{2}/3\|\qb\|$(right). Principal eigenvectors of minimum energy configurations $\qb(\varphi^*)$ for $\beta=0$(left): $\varphi^* = \pi/4$. $\beta=\pm\sqrt{2}/3\|\qb\|$(right): $\varphi^* = \pi/2$. [C] Contributions of integral distortion energy for angle $\varphi$. (top) $\| \nabla \qb \|^2$ with minima at $\varphi = \pi/4,\, 3/4\pi$, (bottom) $-3 \beta L \meanCurvature \langle \shapeOperator,\qb \rangle$ for $\beta = -\sqrt{2}/3\| \qb\|$ with minima at $\varphi = \pi/2$}
	\label{fig:distEnergyQsurfPhi}
\end{figure*}

\begin{figure*}[ht!]
	\begin{center}
				\includegraphics[width=0.95\linewidth]{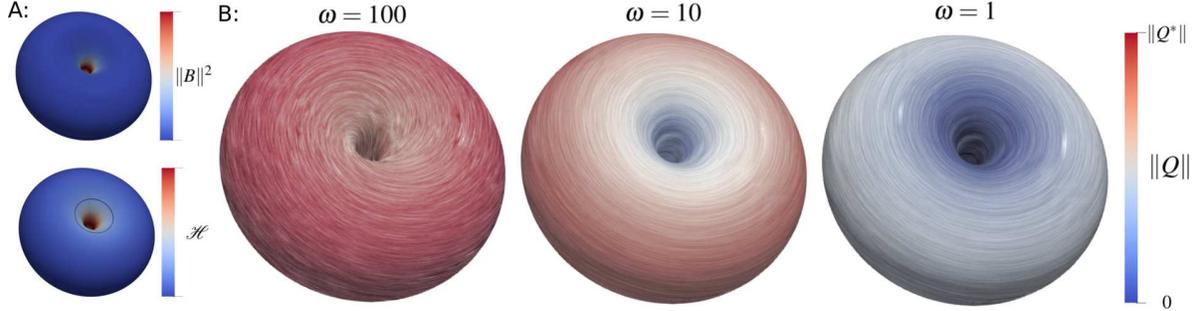}
	\end{center}
	\caption{\textbf{Curvature Impact on Equilibrium Configurations on Thick Torus:} [A] Curvature of thick torus ($R=0.55, r=0.45$). (top) $\| \shapeOperator \|^2 \in [4.7,155]$, (bottom) $\meanCurvature$ changes sign, location indicated by black line, across geometry such that strong positive at center and mild negative at rim $\meanCurvature \in[-3.2,10]$. [B] Equilibrium configurations depending on choice of $\omega$ [$\omega=100$ (left), $\omega=10$ (mid) and $\omega=1$ (right)]. Lines indicate direction of principal director of $\Qb$, obtained by LIC. Colors denote $\| \Qb\| \in [0, \|\Qb^*\|]$, where $\| \Qb^*\|$ is given by  minima value of state potential ($a=1/4,\, b=-4,\, c=1$), (blue) isotropic, (red) nematic. Simulations are performed with $L=1$, $\omega_{\beta}=10$ and $\omega_{\normal}=10^3$.}
	\label{fig:distStateEnergyThickTorus}%Inital values are prescribed as circumferential prinicipal director field and $\| \Qb\| = \|\Qb^* \|$.
\end{figure*}

Fig. \ref{fig:betaModelPhaseCoexistence} further highlights the qualitative differences of the three models on a torus (with two major radii $r,R$), which is chosen as a prototypical surface with varying curvature, which avoids the presence of topological defects.

\section{Impact of Curvature in Surface Models}
\label{sec:CurvedThinFilm}

To investigate the impact of $\beta$ on the geometry coupling mechanisms and to demonstrate the sketched effects we consider two numerical experiments.

\paragraph*{Distortion Energy Minima} To focus on the first coupling term in the surface Landau-de Gennes Q-tensor energy \eqref{eq:CurvedSurfaceBetaExplicitEnergy}, $6 \meanCurvature \beta \langle \shapeOperator, \qb \rangle$, we consider a uniform director field $\dir(\varphi) =  \cos(\varphi) \dirSurf_r - \sin(\varphi) \dirSurf_R$ on a torus defined by linear combination of two director modes $\dirSurf_r$ and $\dirSurf_R$ as shown in Fig. \ref{fig:distEnergyQsurfPhi}-A. With this director field we define $\qb(\varphi) = (\dirSurf(\varphi)\otimes\dirSurf(\varphi) - 1/2\gb)$ with fixed norm. For given $\beta$ we can now variate $\varphi$ and evaluate the distortion energy contributions. In this set up the isotropic coupling term is constant as well as $\|\nabla \beta \|^2$, furthermore the models $\beta = -1/3S^*$ and $\beta = \pm\sqrt{2}/3\|\qb\|$ coincide. For $\beta = 0$ the directed coupling term vanishes and the minimum of distortion energy is defined by the $\| \nabla \qb \|^2$ contribution. As shown in Fig. \ref{fig:distEnergyQsurfPhi}-C(top) two minima exist. The minimum $\varphi^* = \pi/4$ results in a surprising $\qb$ configuration, shown in Fig. \ref{fig:distEnergyQsurfPhi}-B(left). The second minimum $\varphi^*=3/4 \pi$ yields a corresponding configuration with in-plane $90^{\circ}$ rotated eigenvectors. In the model $\beta = \pm\sqrt{2}/3\|\qb\|$ and on the chosen geometry the $-3 \beta L \meanCurvature \langle \shapeOperator,\qb \rangle$ contribution dominates the distortion energy, see Fig. \ref{fig:distEnergyQsurfPhi}-C(bottom) such that the minimum is achieved for $\varphi^*=\pi/2$. The corresponding Q-tensor configuration consist of principal eigenvectors aligned with lines of minimal curvature, see Fig. \ref{fig:distEnergyQsurfPhi}-B(right).

Reviewing these results we conclude, once more, that degenerate ($\beta=0$) surface Landau-de Gennes Q-tensor models describe a substantial different type of physical systems than surface models with non-degenerate Q-tensors. Retaining the 3D nature of liquid crystals in the surface model yields an intensified geometry-ordering coupling.

\paragraph*{Balancing Curvature Effects for $\beta = \pm\sqrt{2}/3\|\qb\|$} To obtain an intuition on the possible interactions of the geometry-order couplings we consider a thick torus ($R=0.55$, $r=0.45$) where $\meanCurvature$ changes sign, see Fig. \ref{fig:distStateEnergyThickTorus}-A. On this surface we evaluate the equilibrium states $\Qb$ of the $\beta = \pm\sqrt{2}/3\|\qb\|$ model for several values of $\omega$, see Fig. \ref{fig:distStateEnergyThickTorus}-B. Recalling $\omega$ as factor weighting the distortion contribution versus the state potential we observe for $\omega=100$ a dominant state potential such that a uniform nematic phase, $\|\Qb\| = \|\Qb^*\|$, as prescribed by the state potential, is enforced across the entire surface, suppressing the effects of the isotropic geometry coupling. In contrast, the directed geometry coupling is not affected and we observe a strong forcing towards the -geometry induced- preferred alignment such that a non uniform ordering of principal directors is yielded, see Fig. \ref{fig:distStateEnergyThickTorus}-B(left). For regions of positive $\meanCurvature$ the alignment is with minimal curvature lines (outer part), whereas for negative $\meanCurvature$ the alignment is with maximal curvature lines (inner part). Similar effects have also been reported for polar liquid crystals within a surface Frank-Oseen model \cite{Segattietal_M3AS_2016,Nestleretal_JNS_2018}. Weakening the state potential by choosing $\omega=10$ we observe a curvature induced phase transition with a localized isotropic phase at the inner part of the torus, where $\|\shapeOperator\|^2$ has its maximum. This area coincides with the area of strongest directed geometry coupling. Finally matching the distortion contribution and state potential by $\omega=1$ we yield a global isotropic phase since the isotropic coupling distorts the state potential such that only the isotropic $\Qb=0$ phase remains stable.

The isotropic geometry coupling term, $\| \shapeOperator\| ^2 ( \trace \qb^2 + \frac{9}{2}\beta^2)$, turns out to be the dominating effect compared to the directed geometry coupling term $6 \meanCurvature \beta \langle \shapeOperator, \qb \rangle$. Only in situations where the state potential is strong enough to suppress the isotropic coupling the effects of directed coupling become traceable. In this situation the geometry induces a preferred alignment, see also \cite{nitschke2018nematic} for similar results on ellipsoidal geometries. For geometries with strong variations in $\meanCurvature$, including sign changes, this leads to nonuniform ordering. An effect obviously possible only in models with $\beta \neq 0$. 

\section{Discussion and Conclusions}
\label{sec:Conclusion}
Exploring a thin film limit of the 3D Landau-de Gennes Q-tensor model allows to bridge the gap to previously proposed surface models. Planar anchoring at the boundary of the thin film was thereby used to fix the boundary normal as eigenvector with eigenvalue $\beta$, which motivated the decomposition of the tensorial order parameter $\Qb$ in tangential $\qb$ and normal $\beta$ parts. We will now review the central results of the models with different choices of $\beta$, discuss suitable application scenarios, assess their inherent couplings to the curvature of the surface and discuss possible experiments to confirm these results. 

\paragraph*{Discussion}
The first class of surface Landau-de Gennes Q-tensor models considers $\beta = 0$ and is usually labeled as planar degenerate Q-tensors \cite{Kraljetal_SM_2011,Napolietal_PRE_2012,Napolietal_PRL_2012,Jeseneketal_SM_2015}. We observed the models to exhibit essentially 2D characteristics like eigenvalue spectra of $\Qb$ matching 2D mean field theory. These models also show a continuous/second order isotropic-to-nematic phase transition, which contradicts results on the existence of first-order isotropic-to-nematic phase transitions in 2D \cite{vanEnteretal_PRL_2002,vanEnteretal_JPA_2006}. Concerning the coupling with geometric properties the model in \cite{Kraljetal_SM_2011} did only account for intrinsic curvature effects. Additional extrinsic contributions, as proposed in 
\cite{Mbangaetal_PRL_2012}, have been considered in \cite{Napolietal_PRE_2012,Napolietal_PRL_2012,Jeseneketal_SM_2015}. However, they only account for a weak isotropic geometry coupling. Prefered alignment of the director field with the principle curvature lines is not present. 

For the second class, where $\beta \neq 0$, the obtained models retain the characteristics of the 3D Landau-de Gennes Q-tensor model. For unconstrained values of $\beta$, eqs. \eqref{eq:q} and \eqref{eq:beta}, the uniaxiality can be perturbed by curvature. Using a 3D biaxiality measure fixes $\beta = \pm\sqrt{2}/3\|\qb\|$ and enforces uniaxial $\Qb$. Further investigations confirmed that for uniaxial Q-tensors the fundamental properties of eigenvalue spectra, first order isotropic-to-nematic phase transition and phase coexistence stay preserved. The model $\beta = -1/3 S^*$, as proposed in \cite{nitschke2018nematic}, has proven to reproduce the 3D characteristics, under the assumption of $\|\Qb\| = \|\Qb^*\|$, except for phase coexistence. It can be considered as a simplified surface Landau-de Gennes Q-tensor model suitable for uniaxial nematic liquid crystals far from phase transition temperature. Furthermore $\beta \neq 0$ introduces additional curvature coupling terms. In addition to isotropic geometry coupling, alignment of the director field with the principle curvature lines is considered. Parameter studies for a torus indicate that, depending on the strength of the curvature, phase transitions can be enforced, leading to phase coexistence and locally confined isotropic regions within a nematic phase or even an uniformaly isotropic phase if the curvature effect is strong enough.

These discrepancies in response to curvature of the $\beta=0$ and $\beta \neq 0$ models provide a motivation for in-vitro experiments to assess the prevalent 2D or 3D nature of liquid crystal which are confined to curved surfaces. 

\paragraph*{Conclusion}
With the presented derivations and arguments we have provided a comprehensive study unifying recent approaches for surface Landau-de Gennes Q-tensor theories for uniaxial nematic liquid crystals confined to curved surfaces. By introducing a new surface parameter
$\beta$ we have classified different surface limits for nematic phases on curved manifolds. Essentially, $\beta$
measures the ability of the orientational degrees of freedom to fluctuate in the direction perpendicular to the curved
surface while the  particle centers are constrained on the manifold in the thin-film limit. In terms of physics,
these are the imposed anchoring conditions at the surface. In particular,
we have identified two classes of models as special limits which could be related to liquid crystals with prevalent 2D or 3D characteristics. The distinct response to curvature of this two model classes enable a path to determine the proper models for the liquid crystal systems by in-vitro experiments. For the future it would be interesting to link the new classes of surface Landau-de Gennes Q-tensor models
studied in this paper to particle-resolved models where anisotropic apolar particles are bound to curved interfaces.
It remains to be understood how different anchoring conditions of the particles at the surface
can be mapped and described effectively by our coarse-grained mean-field-like approach.
In principle, varying the anchoring conditions should result in different coupling parameters used
in our surface free energy \eqref{eq:anchoring}.
In particle-resolved computer simulations, different anchoring conditions can just be implemented
by an explicit orientational coupling to the curved interface. In actual
experiments on colloids bound to curved interfaces (see e.g.\ \cite{Stebe}), on Pickering emulsion droplets
\cite{Lou}  the anchoring conditions can be conveniently changed by the pH \cite{Qin} or by
changing the thermodynamic parameters.

\section*{Conflicts of interest}
There are no conflicts to declare.

\section*{Acknowledgments}
AV acknowledges financial support from DFG through VO899/19. HL acknowledges financial support from DFG through LO 418/20. We further acknowledge computing resources provided by JSC under grant HDR06 and ZIH/TUDresden.
%%%END OF MAIN TEXT%%%

%If notes are included in your references you can change the title from 'References' to 'Notes and references' using the following command:
%\renewcommand\refname{Notes and references}
\renewcommand\refname{References}
%%%REFERENCES%%%
%\bibliography{short_names,rsc}
%\bibliographystyle{plain} %the RSC's .bst file

%======
\appendix
\section{Appendix}

\subsection{Details on Uniaxiality Condition}
To obtain the surface uniaxiality condition \eqref{eq:surfaceUniaxiality} we can insert the Q-tensor decomposition \eqref{eq:qtensorDecomposition} into the biaxiality measure \eqref{eq:BulkEnergyOneConst}. Another elegant approach is to consider the eigenvalue spectra \eqref{eq:EigenvalueSpectraQbeta} and insert it into the biaxiality measure: $Q\mbox{ uniaxial} \Leftrightarrow U(Q) = 0 \Leftrightarrow 6 (\traceR \Qb^3)^2 = (\traceR \Qb^2)^3$. 
We obtain $\traceR \Qb^2 = \frac{1}{2} S^2 + \frac{3}{2} \beta^2$ and $\traceR \Qb^3 = - \frac{3}{4} \beta S^2 + \frac{3}{4} \beta^3$ such that the uniaxiality condition $U(Q)=0$ can be expressed by, assuming $S\neq 0$, $\beta^2 \left( 18 S^2 - 81 \beta^2 \right) = S^4$. This holds iff $\beta^2 = 1/9 S^2$ which translates to $\beta = \pm \sqrt{2}/3 \| \qb\|$.

\subsection{Derivation of Thin Film Limit} \label{sec:derivationOfThinFilmLimit}

Thin film limits require a reduction of degrees of freedom.  We deal with this issue by setting Dirichlet boundary conditions for the normal parts of \( \Qb \) and postulate
a priori a minimum of the free energy on the boundary of the thin film. This is achieved by considering natural boundary condition of the weak Euler-Lagrange equation. In this setting we restrict the density of \(  \mathcal{F} \) to the surface and integrate in normal direction to obtain the surface energy \( \mathcal{F}^{\surf} \). We closely follow \cite{nitschke2018nematic}, use the notation introduced there and only point out differences. 

The free energy \eqref{eq:BulkEnergyOneConst} in the thin film \( \surfh \) in index notation reads
\begin{align*}\label{eq:ThinFilmEnergyIndices}
\mathcal{F}[\Qb] &=
\int_{\surfh} 
\frac{L}{2} \! Q_{IJ;K}Q^{IJ;K} \dV \\
& \quad + \omega \int_{\surfh} \!\!\left(
a Q_{IJ}Q^{JI} + \frac{2}{3}b Q_{IJ}Q^{JK}\tensor{Q}{_{K}^{I}} + c Q_{IJ}Q^{JK}Q_{KL}Q^{LI}\right)
\dV. 
\end{align*}
For the choice of essential boundary conditions, we require that \( \Qb \) has to have two eigenvectors in the boundary tangential bundle and the remaining eigenvector has to be the boundary
normal, \ie, for \( \Pb\in\tangent\partial\surfh \) a pure covariant representation of \( \Qb \) at the boundary is
\begin{align}
\Qb = S_{1} \Pb^{\flat}\otimes\Pb^{\flat} + S_{2} \normal^{\flat}\otimes\normal^{\flat} - \frac{1}{3}\left( S_{1}+S_{2} \right)\Gb
\end{align}
with scalar order parameters \( S_{1} \) and \( S_{2} \).
Hence, it holds \( Q_{i\xi} = Q_{\xi i} = 0 \) and \( Q_{\xi\xi}= \frac{1}{3} (2S_{2} - S_{1}) \) at the boundaries.
The remaining boundary conditions are considered as natural boundary conditions
\( 0 = Q_{ij;\xi}\) at \(\partial\surfh\).
According to the normal eigenvalue \(\beta\) of \(\Qb\) at \(\surf\), we extend this scalar field to the thin film \(\surfh\), 
\ie, \( Q_{\xi\xi}= \hat{\beta} \) with \(\hat{\beta}|_\surf = \beta\).

We can relate the anchoring conditions to surface identities by sums and differences of Taylor expansions at the upper and lower boundary, see \cite{nitschke2018nematic}.
This results in
\begin{align*}
\atsurf{Q_{i\xi}} &= \atsurf{Q_{\xi i}} = \landau(h^{2})
&\atsurf{\partial_{\xi}Q_{i \xi}} &= \atsurf{\partial_{\xi}Q_{\xi i}} = \landau(h^{2}) \notag \\
\atsurf{Q_{\xi\xi;\xi}} &= \landau(h^{2}) & \atsurf{Q_{ij;\xi}} &= \landau(h^{2}) \formPeriod \label{eq:AnchoringExpansion}
\end{align*}
The restricted Q-tensor \( \left\{ \atsurf{Q_{ij}} \right\}\in\tangent^{(2)}\surf \) is not a Q-tensor.
We have \( \trace\left\{ \atsurf{Q_{ij}} \right\} = \atsurf{\traceR\Qb} - \atsurf{Q_{\xi\xi}} = -\beta\). To ensure Q-tensor properties we introduce the projection $\ProjQ:\tb \mapsto \frac{1}{2}\left( \tb + \tb^{T} - (\trace\tb)\gb \right) \formComma$
and define
\begin{align}
\qb := \ProjQ\left\{\atsurf{Q_{ij}}\right\} = \left\{\atsurf{Q_{ij}}\right\} + \frac{\beta}{2}\gb.
\end{align}

We can determine all remaining covariant derivatives restricted to the surface by
\begin{align}
\atsurf{Q_{i\xi ;\xi}} \!&=\! \atsurf{Q_{\xi i;\xi}}
\!=\! \atsurf{\partial_{\xi}Q_{i \xi}} \!- \atsurf{\GGamma_{\xi i}^{K}Q_{K\xi}} \!=\! \landau(h^{2}) \notag\\
\atsurf{Q_{\xi\xi;k}} 
\!&=\! \atsurf{\partial_{k}Q_{\xi\xi}} \!- 2\atsurf{\GGamma_{k \xi}^{L}Q_{L\xi}} \!=\! \beta_{|k} + \landau(h^{2})\notag\\
\atsurf{Q_{i\xi ;k}} \!&=\! \atsurf{Q_{\xi i;k}}
\!= \atsurf{\partial_{k}Q_{i \xi}} \!- \atsurf{\GGamma_{k i}^{l}Q_{l\xi}} \!- \atsurf{\GGamma_{k i}^{\xi}Q_{\xi\xi}} \!- \atsurf{\GGamma_{k \xi}^{l}Q_{il}}\notag\\
\!&=\!  \left[\qb\shapeOperator\right]_{ik} - \frac{3}{2}\beta B_{ik} + \landau(h^{2}) \notag\\
\atsurf{Q_{ij;k}}
\!&=\! \atsurf{\partial_{k}Q_{i j}} \!- \atsurf{\GGamma_{ki}^{l}Q_{lj}} \!- \atsurf{\GGamma_{ki}^{\xi}Q_{\xi j}}
\!- \atsurf{\GGamma_{kj}^{l}Q_{il}} \!- \atsurf{\GGamma_{kj}^{\xi}Q_{i \xi}} \notag\\
\!&=\!\atsurf{\partial_{k}Q_{i j}} \!- \atsurf{\Gamma_{ki}^{l}Q_{lj}} \!- \atsurf{\Gamma_{kj}^{l}Q_{il}} + \landau(h^{2})\label{eq:thinfilmgradienttosurface}\\
\!&=\! \left( q_{ij} - \frac{\beta}{2} g_{ij} \right)_{|k} \!+ \landau(h^{2})
= q_{ij|k} - \frac{1}{2}\beta_{|k}g_{ij} +  \landau(h^{2}) \formPeriod \notag
\end{align}
The contributions of the energy density read  
\begin{align}
\atsurf{\left\|\nabla\Qb\right\|^2_{\Gb}} 
&= \atsurf{Q_{ij;k}Q^{ij;k} + 2Q_{i\xi;k}Q^{i\xi;k} + 2Q_{\xi\xi;k}Q^{\xi\xi;k}} + \landau(h^{2})\notag\\
&= \left\|\nabla\gb\right\|^2_{\gb} + \frac{3}{2}\left\|\nabla\beta\right\|^2_{\gb} + 2\left\|\qb\shapeOperator-\frac{3}{2}\beta\shapeOperator\right\|^2_{\gb}
+\landau(h^{2})\notag\\
\atsurf{\traceR\Qb^{2}}
&= \trace\qb^{2} + \frac{3}{2}\beta^{2} +\landau(h^{2})\notag\\
\atsurf{\traceR\Qb^{3}} 
&= \frac{3}{2}\beta\left( \frac{\beta^{2}}{2} - \trace\qb^{2} \right) +\landau(h^{2}) \label{eq:energysurfBulkQ3}\\
\atsurf{\traceR\Qb^{4}} 
&= \trace\qb^{4} + \frac{3}{2}\beta^{2}\trace\qb^{2} + \frac{9}{8}\beta^{4} +\landau(h^{2})\formPeriod \notag      
\end{align}
Using
\begin{align}\label{eq:qwertz}
2\left\|\qb\shapeOperator-\frac{3}{2}\beta\shapeOperator\right\|^2
&= \left\|\shapeOperator\right\|\left(\trace\qb^2 + \frac{9}{2}\beta^2\right) - 6\meanCurvature\beta\left<\shapeOperator,\qb\right>
\end{align}
which follows from \cite{nitschke2018nematic}(Corollary A.4.), adding all contributions up and denoting the free energy densities by $F$ and $F^\surf$, we obtain for \( h\rightarrow 0 \)
\begin{align}\label{eq:rectangleEnergie}
\frac{1}{h}\mathcal{F} &= \frac{1}{h}\int_{\surfh}F\dV
= \frac{1}{h}\int_{-\frac{h}{2}}^{\frac{h}{2}} \int_{\surf}  \left( 1- \xi\meanCurvature + \xi^{2}\gaussianCurvature \right)F\dS\dxi\notag\\
&= \int_{\surf} F^{\surf} \dS + \landau(h^{2})
= \mathcal{F}^{\surf} + \landau(h^{2}) 
\longrightarrow \mathcal{F}^{\surf} \formPeriod
\end{align}

\subsection{Variational Derivatives} \label{sec:variationalDerivatives}

The first variation of 
\begin{align}
\energy^{\surf}[\qb,\beta] = & \frac{1}{2}\int_{\surf} L \| \nabla \qb \|^2 + \frac{3}{2}L \| \nabla \beta \|^2  \nonumber \\ 
& \quad\quad- 6 L \meanCurvature \beta \langle \shapeOperator, \qb \rangle  +  L\| \shapeOperator\| ^2 \left( \trace \qb^2 + \frac{9}{2}\beta^2\right)\,\mathrm{d}\surf \nonumber \\
&+ \omega \int_{\surf}\frac{1}{2}\left(2a -2b\beta + 3c\beta^2\right) \trace \qb^2 + c \trace \qb^4 \nonumber \\
& \quad\quad + \frac{\beta^2}{8}\left(12a +4b\beta +9 c \beta^2\right)  \,\mathrm{d}\surf \formComma
\end{align}
\wrt\ surface Q-tensor \(\psi\) and scalar \(\varphi\) perturbations for \(\qb\) and \(\beta\), reads
\begin{align*}
\delta \mathcal{F}^{\surf}
&=  L\int_\surf  \left\langle \nabla\qb, \nabla\psib \right\rangle
+ \left\langle \| \shapeOperator\| ^2\qb - 3\meanCurvature\beta\ProjQ\shapeOperator , \psib  \right\rangle \notag\\
&\quad\quad\quad + \frac{3}{2}\left<\nabla\beta, \nabla\varphi\right> - 3\left( \meanc\left\langle \shapeOperator, \qb\right\rangle 
- \frac{3}{2}\left\|\shapeOperator\right\|^2\beta\right)\varphi
\,\dS\\
&\quad + \omega \int_\surf (2a - 2b\beta + 3c\beta^2)\left\langle \qb, \psib \right\rangle
+ 2c\trace\qb^{2}\left\langle \qb, \psib \right\rangle \\
&\quad\quad\quad + \left(\frac{9}{2}c\beta^3 + \frac{3}{2}\beta^2 + 3\left(c\trace\qb^2+ a\right)\beta - b\trace\qb^2\right)\varphi \,\dS \formComma
\end{align*}
where \(\beta\) is independent.
If \(\beta\) is constant, or prescribed generally, the associated perturbation \(\varphi\) vanish.
To enforce uniaxiality \eqref{eq:surfaceUniaxiality}, we add the penalty energy
\begin{align*}
\mathcal{F}^{\surf}_{\text{uni}}
&= \omega_\beta \int_{\surf} \frac{1}{4} \left( \beta^2 - \frac{2}{9} \| \qb \|^2 \right)^2\,\mathrm{d}\surf
\end{align*}
to \(\mathcal{F}^{\surf}\), with \(\omega_\beta > 0\). Its first variation reads
\begin{align*}
\delta\mathcal{F}^{\surf}_{\text{uni}}
&= \omega_\beta \!\int_\surf \!\!- \frac{2}{9} \left( \beta ^2 - \frac{2}{9} \trace\qb^2 \right) \langle \qb,\psib \rangle + \left( \beta^2 - \frac{2}{9} \trace\qb^2 \right) \beta \varphi \,\dS \!\formPeriod
\end{align*}
From this, the $L^2$-gradient flows lead to the consider evolution equations for $\qb$ and $\beta$.

\subsection{$L^2$-Gradient Flows} \label{sec:L2Gradientflow}

Instead of relating the energies $\mathcal{F}$ and $\mathcal{F}^\surf$, also the evolution equations in $\surfh$ and on $\surf$ can be related. 
Similar calculations as in \eqref{sec:derivationOfThinFilmLimit} give
\begin{align*}
\atsurf{\left\langle \nabla_{L^{2}}\mathcal{F}^{\surf_h}, \Psib \right\rangle}
&= \left\langle \nabla_{L^{2}}^{\qb}\mathcal{F}^{\surf}, \psib \right\rangle +  \left\langle \nabla_{L^{2}}^{\beta}\mathcal{F}^{\surf}, \varphi \right\rangle + \landau(h^{2}),
\end{align*}
with appropriate bulk and surface Q-tensors $\Psib$ and $\psib$ and scalar $\varphi$ perturbations, \st\ 
\(\atsurf{\Psi_{\xi\xi}}=\varphi\).
This condition is the principal difference to the calculations in \cite{nitschke2018nematic}.
Note that we used \eqref{eq:qwertz} in the derivation, weakly in Q-tensor direction \(\psib\), \ie
\begin{align*}
\int_\surf \left\langle \qb\shapeOperator - \frac{3}{2}\beta\shapeOperator, \psib\shapeOperator \right\rangle\,\dS
&=\int_\surf  \left\langle \| \shapeOperator\| ^2\qb - 3\meanCurvature\beta\ProjQ\shapeOperator , \psib  \right\rangle \,\dS \formPeriod
\end{align*}
Moreover, as \( \partial_{t}\gb = 0 \) for a stationary surface, we obtain 
\(\atsurf{\left<\partial_t\Qb,\Psib\right>_{\Gb}} = \left<\partial_t\qb,\psib\right>_{\gb} + \frac{3}{2}(\partial_t\beta)\phi + \landau(h^{2})\) and as in \eqref{eq:rectangleEnergie}
\begin{align}\label{eq:FirstVarThinVSSurf}
&\frac{1}{h}\int_{\surfh}\left\langle \partial_{t}\Qb + \nabla_{L^{2}}\mathcal{F}, \Psib \right\rangle\dV\\
&\quad =  \int_{\surf}\left\langle \partial_{t}\qb + \nabla_{L^{2}}^{\qb}\mathcal{F}^{\surf}, \psib \right\rangle
+\left(\frac{3}{2}\partial_t\beta + \nabla_{L^{2}}^{\beta}\mathcal{F}^{\surf}\right)\varphi \dS + \landau(h^{2})\formPeriod
\end{align}
Therefore, the relaxation velocity parameter, for the \(L^2\)-gradient flow \wrt\ \(\beta\), has to be \(\frac{2}{3}\) for consistency of the the time-depending thin film and surface problems \wrt\ \(h\).
The surface evolution equations thus read
\begin{align}
\partial_t \qb &= L \Laplace^{DG}_\surf \qb - ( L(\meanCurvature^2 - 2 \gaussianCurvature) - \omega(2 a -2b \beta + c(3 \beta^2 + 2 \trace \qb^2)))\qb \nonumber \\
& \quad + 3 L \meanCurvature \beta \left(\shapeOperator - \frac{1}{2} \meanCurvature \gb \right) + \omega_\beta \frac{2}{9}\left[ \beta^2 - \frac{2}{9}\|\qb\|^2 \right]\qb \formComma \label{eq:qPenalty} \\
\!\!\!\partial_t \beta &= L \laplace_\surf \beta - \omega (3c \beta^3 + b \beta^2) - \omega(2a + 2c \trace \qb^2) \beta + 2 L \meanCurvature \langle \shapeOperator, \qb \rangle \nonumber \\
& \quad  - 3L(\meanCurvature^2 - 2 \gaussianCurvature) \beta + \frac{2}{3} b \trace \qb^2 -\omega_\beta\frac{2}{3}\left[ \beta^2 - \frac{2}{9}\|\qb\|^2\right]\beta \formPeriod\label{eq:betaPenalty}
\end{align}

\subsection{Numeric Solution Procedure}
To numerically solve the tensor- and scalar-valued surface PDEs \eqref{eq:qPenalty} and \eqref{eq:betaPenalty}, we use the surface FEM approaches of \cite{Nestleretal_JCP_2019} and \cite{dziuk2013finite}, respectively. The approach in \cite{Nestleretal_JCP_2019} extends previous ideas for vector-valued surface PDEs \cite{Nestleretal_JNS_2018,Jankuhnetal_IFB_2018,Hansboetal_IMA_2020} to tensors of arbitrary degree. The idea of these approaches is to reformulate the problems in Cartesian coordinates and to penalize normal components. This allows for a componentwise solution using tools for scalar-valued surface PDEs, e.g. \cite{dziuk2013finite}. The penalty term, added to $\mathcal{F}^\surf$ reads $\omega_{\normal} \int_{\surf} \frac{1}{2}\| \qb \cdot \normal \|^2 \,\mathrm{d}\surf$ with $\omega_{\normal} > 0$. This leads to an additional term in \eqref{eq:qPenalty} reading $\omega_{\normal} (\normal\otimes\normal \cdot \qb)$. 

To address the nonlinearity of the system of PDEs we consider a Newton method. We solve the temporal discretized problem as a sequence of time steps $[\hat{\qb},\hat{\beta}] \rightarrow \qb,\beta$, where $\partial_t \qb \approx (\qb - \hat{\qb})/\tau$ and $\partial_t \beta \approx (\beta - \hat{\beta})/\tau$. We denote with $L\mathcal{Q}$, $L\mathcal{B}$ and $N\mathcal{Q}$, $N\mathcal{B}$ the collections of linear and nonlinear operators of the time step problems, where $\mathcal{Q}$ and $\mathcal{B}$ refers to the $\qb$ and $\beta$ state equation. The single Newton iteration $k \rightarrow k+1$ reads
\begin{align*}
&\delta_{\qb} N\mathcal{Q}(\qb^k,\beta^k)[\qb^{k+1}] + \delta_{\beta} N\mathcal{Q}(\qb^k,\beta^k)[\beta^{k+1}] + L\mathcal{Q}(\qb^{k+1},\beta^{k+1})\nonumber \\
=& \delta_{\qb} N\mathcal{Q}(\qb^k,\beta^k)[\qb^k] + \delta_{\beta} N\mathcal{Q}(\qb^k,\beta^k)[\beta^k] - N\mathcal{Q}(\qb^k,\beta^k)\\
& \delta_{\qb} N\mathcal{B}(\qb^k,\beta^k)[\qb^{k+1}] + \delta_{\beta} N\mathcal{B}(\qb^k,\beta^k)[\beta^{k+1}] + 
L\mathcal{B}(\qb^{k+1},\beta^{k+1}) \nonumber \\
=& \delta_{\qb} N\mathcal{B}(\qb^k,\beta^k)[\qb^k] + \delta_{\beta} N\mathcal{B}(\qb^k,\beta^k)[\beta^k] - 
N\mathcal{B}(\qb^k,\beta^k)
\end{align*}
To evaluate $[\qb,\beta]$ we solve the Newton iterations until 
\begin{align*}
\left(\int_{\surf}\| \qb^k - \qb^{k+1}\|^2 + | \beta^k - \beta^{k+1}\|^2  \,\mathrm{d}\surf \right)^{1/2} < \theta.
\end{align*}
The resulting linear surface PDE's are solved by the surface FEM methods \cite{Nestleretal_JCP_2019,dziuk2013finite} which are implemented in the adaptive FEM toolbox AMDiS \cite{vey2007amdis,witkowski}.

To asses the quality of enforcing tangentiality and uniaxiality by the introduced penalty terms, simulations are performed with $\omega_{\normal} \in [10^1, 10^3]$ and $\omega_{\beta} \in [10^1, 10^3]$ on a torus ($R=2,\, r=0.5$) for $\omega=10$, $a=1/4$, $b=-4$ and $c=1$. Across the studied parameters we obtain for tangential alignment $\|\qb \cdot \normal\| < 10^{-3}$ and for uniaxiality $|\beta^2+\sqrt{2}/3\|\qb\|^2|/\|\Qb^*\| < 10^{-3}$ at each point of $\surf$.

\end{document}